# A New Indicator of Nonlinear Gravitational Clustering

J.S.Bagla[1]

Inter-University Centre for Astronomy and Astrophysics, Post Bag 4, Ganeshkhind, Pune - 411 007, INDIA.

**Abstract.** Alignment of velocity and acceleration before shell crossing, and later misalignment are used to define velocity contrast, an indicator of dynamical state of matter undergoing gravitational collapse. We use this to study bias in clustering properties of dynamically nonlinear mass.

## 1. Introduction

The problem of gravitational dynamics in an expanding universe can be reduced to the corresponding Newtonian problem for a nonrelativistic collisionless medium, if scales of interest are small compared to the hubble radius. Further if $\Omega_o = \Omega_{matter} = 1$ then following equations describe the system.

$$\frac{d\mathbf{u}}{da} = -\frac{3}{2a}(\mathbf{u} - \mathbf{g}) \quad ; \quad \nabla^2 \psi = \frac{\delta}{a},$$
$$\mathbf{g} \equiv -\nabla \psi \equiv -\frac{2}{3H_0^2}\nabla\varphi \quad ; \quad \mathbf{v} \equiv a\dot{\mathbf{x}} \equiv a\dot{a}\mathbf{u} \qquad (1)$$

Here $\mathbf{u} = d\mathbf{x}/da$ is peculiar velocity, $\mathbf{g}$ is "acceleration" and $\psi$ is gravitational potential. These quantities have been defined in a dynamically relevant manner, their relation with the usual definitions are given in (1).

All concepts introduced here can be defined for a density and velocity field or its particle realization. In this paper, we use particle picture for convenience.

Dynamical evolution described by (1) can be divided into four regimes. To motivate the form of new indicator, we describe evolution of trajectories in these regimes. In *linear* regime density contrast grows in proportion to scale factor. Peculiar velocities are constant and equal the acceleration, $\mathbf{u}(\mathbf{x}, a) = \mathbf{g}(\mathbf{x}, a)$.

This equality can be extended, approximately, beyond linear regime. Trajectories closely follow those predicted by Zeldovich approximation ( Zeldovich, 1970) upto shell crossing. Particles move with constant peculiar velocity, velocity being fixed at the initial position : $\mathbf{u}(\mathbf{x}, a) = \mathbf{g}(\mathbf{x}_{in}, a_{in})$. Accuracy of this approximation implies $\mathbf{u}(\mathbf{x}, a) \simeq \mathbf{g}(\mathbf{x}, a)$ for the *Zeldovich* regime.

In *quasilinear* regime particles oscillate about pancakes while drifting towards deep potential wells. Alignment of velocity and acceleration is disturbed, except in direction of bulk motion. In *nonlinear* regime, velocities are random and there is no alignment between velocities and acceleration.

---

[1] email : jasjeet@iucaa.ernet.in

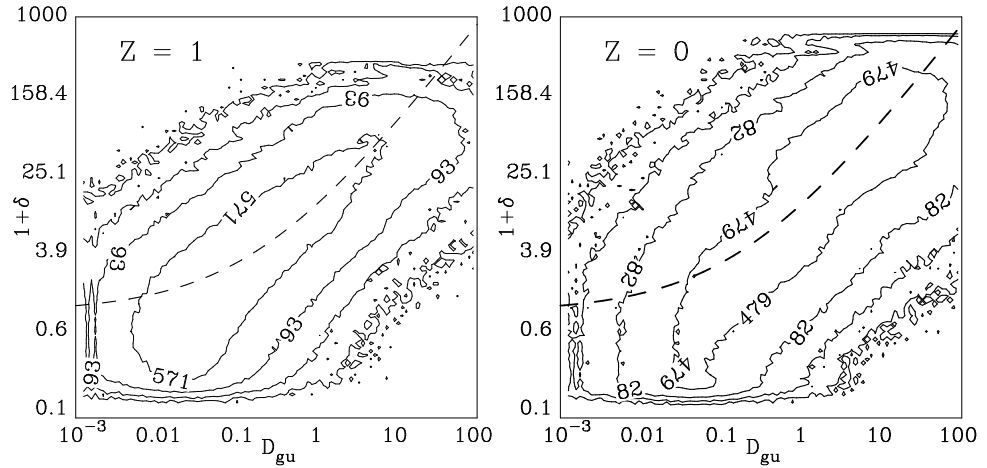

Figure 1. Contours of equal population on $D_{\bf gu}-(1+\delta)$ plane. These have been drawn for CDM at $Z = 1, 0$. Dashed line shows the relation for spherical top hat collapse.

## 2. Velocity Contrast

Above discussion shows that velocity and acceleration are aligned before shell crossing, after this misalignment increases till velocities are randomized. Therefore misalignment of velocity and acceleration [ in direction *and magnitude*] is an indicator of nonlinearity. Velocity contrast quantifies this mismatch as

$$D_{\bf gu} = \frac{({\bf u}-{\bf g})^2}{{\bf u}.{\bf u}} \qquad (2)$$

A comparison of velocity and density contrasts for CDM simulations is presented here. [ Shell crossing is necessary for formation of dynamically nonlinear objects, even if they form by merger of smaller virialised objects. This justifies use of velocity contrast for CDM spectrum.] This simulation used a PM code with $128^3$ particles in a $128^3$ box, with $M_{particle} \simeq 2 \times 10^{11} M_\odot$, so results apply to a CDM universe smoothed at that [ mass] scale.

Figure 1 shows contours of equal population on $D_{\bf gu} - (1+\delta)$ plane for two epochs [ $Z = 1, 0$]. Relation for spherical top hat model has also been plotted for comparison. Average relation between density and velocity contrast coincides with that expected from STH over a large range in both panels. At large values of velocity contrast [ $D_{\bf gu} \sim 50$], there is an upward shift of contours from $Z = 1$ to $Z = 0$. At a given level of nonlinearity in dynamics, say at turnaround, density is highest for spherical and lowest for planar collapse. Therefore an upward shift in the relation implies a shift from predominantly planar collapse towards spherical collapse. Collapsing objects with different shapes contribute to the large dispersion. At low velocity contrast, dispersion arising from this effect is almost an order of magnitude [ See Bagla and Padmanabhan (1995)].

Comparison of velocity and density contrasts for a rich cluster is shown as contours of equal projected density and velocity contrast for a cube in fig.2. There are distinct differences in two indicators, density contours always pick out regions that appear prominent to the eye. Velocity contrast does not always follow the same contours, at times there is no contour corresponding to a pancake. This happens if matter has not shell crossed and is only falling along a "pancake".

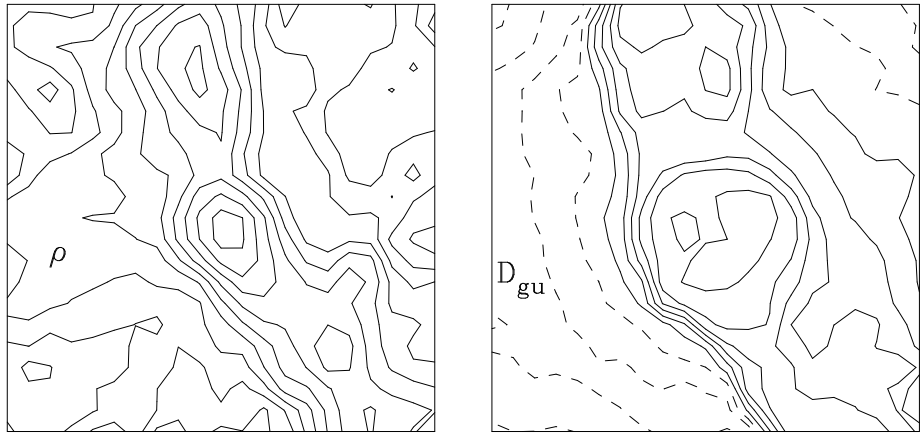

Figure 2. Contours of projected density and velocity contrast for a cube of size $14h^{-1}Mpc$ from a CDM simulation.

Velocity contrast indicates substructure in some density peaks, this results from recent merger of smaller virialised objects or accretion of matter with little nonlinear dynamics. Some dense objects are not visible in contours of velocity contrast. In such objects, bulk motion towards neighboouring density peaks dominates over random velocities within. Highest contour levels for the two indicators do not always coincide, which could arise from presence of infalling material that has not undergone sufficient mixing.

Virialised structures, like galaxies can form only after shell crossing and subsequent mixing of matter. Therefore only regions with sufficiently nonlinear dynamics can host luminous structures. Clustering properties of these regions and underlying mass distribution can be used to define an effective bias :

$$b^2(x) = \frac{J_3^{nonlin}(x)}{J_3^{total}(x)} \quad ; \quad J_3(x) = \int_0^x y^2 \xi(y) dy \qquad (3)$$

where $\xi$ is the correlation function. In fig.3 we have plotted the effective bias parameter as a function of scale for four threshold values of velocity contrast. These figures show that bias is larger for regions with greater nonlinearity in dynamics, and it approaches unity at large scales.

Ability of $D_{\bf gu}$ to delineate regions with different levels of dynamical nonlinearity can be used for studying environmental effects in galaxy formation. Assuming that different types of galaxies form in regions of different levels of nonlinearity, we can compute relative bias for these. Figure 3, shows bias parameter for three bins of velocity contrast. Bias parameter for mass in regions with $.1 < D_{bfgu} < 1$ is smaller than unity. This however could be an artifact of granularity in simulations. Bias parameter for other bins is greater than unity, shows strong variation at small scales and approaches unity at large scales.

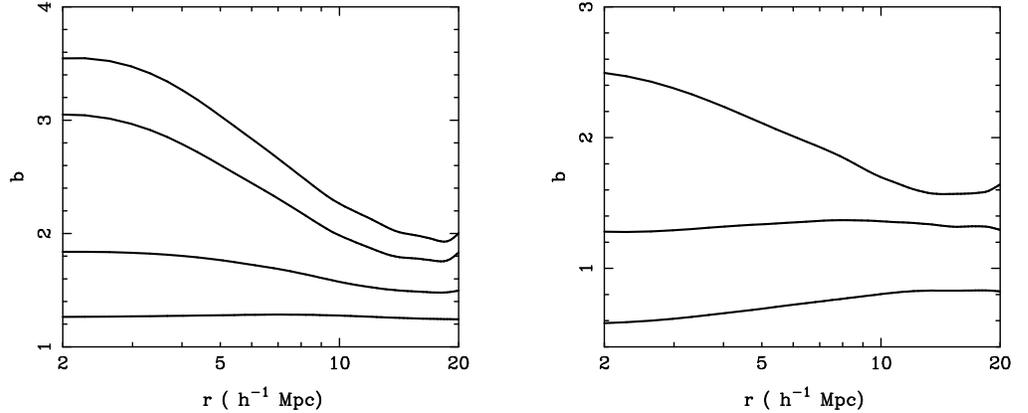

Figure 3. Right panel shows bias for regions with $D_{\mathbf{gu}} > .1, 1, 10$ & 20 respectively. Left panel shows bias for regions in intermediate bins $.1 - 1$, $1 - 10$ & $10 - 20$. Bias is higher for more nonlinear regions in both panels.

## 3. Discussion

Velocity contrast can delineate regions with different levels of nonlinearity in dynamics, this can provide new insight into nonlinear dynamics and clustering. As dynamics of collisional matter is modulated by that of collisionless dark matter, we can use simple modelling to understand some environmental effects. Velocity contrast can also be used as a preliminary filter for computationaly intensive schemes like DENMAX. ( Bertschinger and Gelb 1991).

A comparison of regions selected using velocity contrast with those where luminous structures form in full hydrodynamical simulations can be used for studying influence of dynamics of dark matter on baryonic structures and relative importance of astrophysical processes at various scales.

**Acknowledgments.** I thank T.Padmanabhan for many useful discussions and permission to use some results of work done in collaboration. I thank CSIR India for continued support through Senior Research Fellowship.

## References

Bagla, J.S., & Padmanabhan, T. 1995, IUCAA 9/95, Submitted to MNRAS
Bertschinger,E. & Gelb, J.M. 1991, Computers in Physics, **5**, 164
Zeldovich, Ya.B. 1970, A&A, **5**, 84